\documentstyle[11pt,newpasp,twoside,epsf]{article}
\def\edcomment#1{\iffalse\marginpar{\raggedright\sl#1\/}\else\relax\fi}
\reversemarginpar

\begin{document}
\title{Element stratification in main sequence stars and its effect on stellar oscillations }
 \author{Sylvie Vauclair}
\affil{Laboratoire d'Astrophysique, Observatoire Midi-Pyr\'en\'ees,
14 avenue Edouard Belin, 31400 Toulouse, France}

\begin{abstract}
The element settling due to the combined effects of gravity, 
thermal gradient, radiative acceleration and concentration 
gradient may lead to important abundance variations inside 
the stars, which cannot be neglected in the computations of 
stellar structure. These processes where first introduced to
 account for abundance anomalies in ``peculiar stars", but 
their importance in the so-called ``normal" stars is now 
fully acknowledged, specially after the evidence of helium 
settling in the Sun from helioseismology. These microscopic
 processes work in competition with macroscopic motions, 
like rotation-induced mixing or mass loss, which increase
 the settling time scales. We have recently obtained clear
evidences that asteroseismology of main sequence solar type 
stars can give signatures of the chemical variations inside 
the stars and help for a better understanding of these processes. 
\end{abstract}

\section{ Introduction }

It is clearly admitted nowadays that the element abundances observed in stellar
 outer layers by spectroscopy do not necessarily reflect the chemical 
composition inside the stars. This composition may be altered by microscopic 
diffusion or by accretion, modulated by the macroscopic motions which may occur 
below the outer convective zones, namely rotation-induced mixing, 
internal waves 
or mass motions induced by stellar winds

The importance of element settling inside stars (also called ``microscopic
diffusion") was first recognized as 
an explanation for the 
so-called``chemically peculiar stars" 
(Michaud 1970, Michaud et al. 1976 , Vauclair
et al. 1978a and b). At that time, it was presented as ``the diffusion 
hypothesis", which appeared as a special process added in
some cases to account for stellar abundance peculiarities. 
The fact that element settling is a fundamental process which must occur in
most stars, including the so-called ``normal" stars, has been
proved by helioseismic investigations (Gough et al 1996). It was known for a
 long time that helium and metals should have diffused by
about $20\%$ down from the solar convective zone since the birth of the
Sun up to now (Aller and Chapman 1960). Comparisons between the
sound velocity inside the Sun as computed in the models and as deduced
from the inversion of seismic modes, confirmed that this settling really
occurred.

Element settling is now introduced in most computations of stellar models as a ``standard process". A star is a
self-gravitational gaseous sphere, composed of
a mixture of various gases with different partial pressures, masses and atomic 
spectra. Due to the pressure and thermal gradients and to the selective 
radiative transfer, individual elements diffuse one with respect to the other, 
leading to a slow but effective restructuration.
It is more difficult, in this framework, to understand why the consequences of
diffusion processes are not seen in all the stars than to explain abundance 
anomalies. It is also not easy to account for the fact that, in chemically 
peculiar stars, the observed anomalies are not as strong as predicted by the 
theory of pure microscopic diffusion : these questions are related to the hydrodynamical processes
which occur in stellar radiative zones and compete with element
settling, thereby increasing the time scale of abundance variations.

In the following sections, I will discuss the interaction between 
element settling and stellar oscillations in two parts : 
1) element settling as a stabilizing or an exciting process for triggering
stellar oscillations ; 
2) asteroseismology as a test of element settling inside stars.

\section{Element settling as a stabilizing or an exciting process for triggering stellar oscillations }

The variations of the chemical composition inside the stars, which result from 
element settling, can in some cases lead to the stabilization of layers which 
would otherwise trigger oscillations (e.g. helium depletion in 
Am stars). In 
other cases it can help destabilizing a star which would otherwise be stable
against pulsations (e.g. iron accumulation in some
peculiar stars, or helium diffusion in roAp stars).

\subsection{ Generalities about element settling and radiative accelerations}

Inside the convective regions, the rapid macroscopic motions mix the gas components and force their abundance homogenization. The chemical composition observed in the external regions of cool stars is thus affected by the settling which occurs below the outer convective zones, while in hot stars diffusion occurs directly in the atmosphere, which may lead to abundance gradients or ``clouds" in the spectroscopically observed region. As the settling time scales vary, in first approximation, like the inverse of the density, the expected variations are smaller for cooler stars, which have deeper convective zones. While some elements can see their abundances vary by several orders of magnitude in the hottest Ap stars, the abundance variations in the Sun are not larger than a few $\cong~10\%$. 

While gravitational settling and thermal diffusion lead to a downward motion of all the elements other than hydrogen,
selective radiative transfer may push some elements upwards. This is due to the fact that the elements absorb photons which basically come from the internal layers of the star and re-emit these photons in an isotropic way, thereby gaining a net momentum upwards. This absorption may take place through the continuum (ionization) or through the lines (excitation of the ions). In practice the line process is much more significant than the continuum. 

Radiative accelerations on individual ions strongly depend on their atomic characteristics, which have to be precisely known. They also depend on the sharing of the photon flux with the other elements. It represents the same kind of problems as those encountered for the computations of stellar opacities. This is the reason why the most precise computations presently done on radiative accelerations are the result of collaborations between stellar physicists specialists of diffusion processes and atomic physicists specialists of opacity computations.

The importance of the radiative accelerations on individual elements increase with the effective temperature (Michaud et al 1976). While it is negligible for the Sun (Turcotte et al 1998), it becomes larger than the gravity for most elements in hotter stars (Alecian, Michaud, Tully 1993, Turcotte, Richer and Michaud 1998, Richer et al 1998, Turcotte et al 2000). The Montreal models are presently the only ones which include step by step in the computations the modification of the stellar internal structure induced by the abundance variations. Approximation formulae for the general
computations of radiative accelerations have been derived by Alecian and LeBlanc 2000 and 2002.

The radiative accelerations on the elements vary with depth, according to their ionization stage. It may happen to be smaller than gravity at some depth and larger than gravity below. In this case there is an accumulation of the considered element at that special depth, even if it is depleted in the outer layers. This is the case, for example, for iron inside B, A and F stars (Richard, Michaud, Richer 2001, Richard et al 2002). It is then possible that a new convective zone, due to this iron accumulation, takes place inside the star, thereby changing the way diffusion processes behave. Note that the same phenomenon occurs in horizontal branch stars and is supposed to be the reason for the oscillations of ``SDB stars" (Charpinet et al 1997).

\subsection{Am versus $\displaystyle \delta $-scuti stars }

Among the main sequence stars which lie inside the instability strip, many chemically peculiar stars are found. The magnetic stars will be discussed below. Here I focus on the so-called Am stars, which are found in the H.R. diagram at the same place as the $\displaystyle \delta $-scuti stars. Generally speaking, the former ones show abundance peculiarities, namely a general overabundance of metals (except calcium and scandium), but no oscillations, while the later ones are pulsating but chemically normal. As discussed by Turcotte et al (2000), in A-type stars, almost $70\%$ of non chemically peculiar stars are $\displaystyle \delta $-scuti variables at current levels of sensitivity while most non-variable stars are Am stars. Furthermore, Am stars are slower rotators than $\displaystyle \delta $-scuti stars. This lead Baglin (1972) to suggest a dichotomy between the two kinds of stars. In this region of the H.R. diagram, the stars display two different convective zones in their outer layers : the upper one due to the HI and HeI ionisations and the lower one to the HeII ionisation. The $\displaystyle \delta $-scuti pulsate due to a $\displaystyle \kappa $-mechanism which takes place in the second convective zone. When microscopic diffusion occurs, this convective zone disappears due to helium depletion and the $\kappa $-mechanism cannot take place anymore (Vauclair, Vauclair and Pamjatnikh 1974). 

More recently however, some oscillating Am stars have been discovered (Kurtz 1989, Kurtz et al 1995, Martinez et
al 1999, Joshi et al 2002), which challenge the previously accepted theory. Richer, Michaud and Turcotte (2000) and Turcotte et al (2000) computed models of Am stars in the framework of the Montreal models. They found that, due to the iron accumulation in the radiative zone below the H and He convective zone, a new convective region appears which increases the diffusion time scales compared to the previous models. In these new models, helium is still substantially present in the helium convective zone at the ages of the considered stars. They claim that it is possible to account for the existence of oscillating Am stars close to the cool boundary of the instability strip.
In all these computations, the different convective zones are assumed to be completely connected by overshooting. This assumption, first proposed by
Latour et al 1976 and Toomre et al 1976, is still supported by
recent numerical computations (Toomre, private communication).

\subsection{Rapidly oscillating Ap stars}

Rapidly oscillating Ap stars are very complex objects, due to their magnetic 
structure (Kurtz 2000). The models which have been proposed up to now 
are much too simple to be able to account for all their features. Dolez, Gough and Vauclair (1988) and Vauclair, Dolez and Gough (1991) proposed that the oscillations in roAp stars could be driven by $\displaystyle \kappa $-mechanism in the HeII ionisation zone. As helium is always depleted by diffusion, the idea was that a wind could exist at the magnetic poles, creating a helium overabundance. This model was supposed to be the continuation for cool stars of the model proposed by Vauclair (1975) to account for helium rich stars. In cooler stars, the accumulation should not be visible in the atmosphere, but it should still occur at the place where helium becomes neutral (first ionisation zone). As the $\displaystyle \kappa $-mechanism is driven by the second ionisation zone, the helium accumulation could be efficient only if it was large enough so that its downward wing could appreciably extent down to this region. 

This model for roAp stars has been challenged by Balmforth et al. (2001), who computed the driving of the modes with different helium profiles at the magnetic poles and equator. They found that, contrary to what was assumed before, a helium accumulation induced by diffusion in a wind does not help driving the pulsations. On the contrary, the excitation of the modes is more important when there is less helium in the atmosphere, the basic driving mechanism being induced by hydrogen ionization (this was already suggested by Dziembowski and Goode 1996). Meanwhile the magnetic equator damps the oscillations due to energy loss by turbulence in the remaining convection zone : this could explain why oscillations aligned with the magnetic axis are preferentially excited in these stars. In this respect, recent spectroscopic observations which show evidences of abundance stratification in roAp stars which are not seen in noAp stars may give interesting clues (Gelbmann et al 2000, Ryabchikova et al 2002).

\section{ Asteroseismology as a test of element settling inside stars }

\subsection{The solar case}

Owing to helioseismology, the sound velocity inside the Sun is known with a 
precision 
of $\cong~0.1\%$ and gives evidence of the occurrence of helium settling as 
predicted by diffusion computations. 
Solar models computed in the old ``standard" way, in which the element settling 
is 
totally neglected, do not agree with the inversion of the seismic modes. This 
result 
has been obtained by many authors, in different ways (see Gough et al 1996 and 
references therein). There is a characteristic discrepancy of a few percent, 
just 
below the convective zone, between the sound velocity computed in the models and 
that of the seismic Sun. Introducing the element settling considerably improves 
the consistency with the seismic Sun, but some discrepancies do remain, particularly 
below the convective zone where a peak appears in the sound velocity. 
The reason of this peak is probably due to the steepness of the $\displaystyle 
\mu $-gradient induced
 by pure microscopic diffusion (Richard et al 1996). The helium profiles directly 
obtained from helioseismology (Basu 1997 and 1998, Antia and Chitre 1998) show 
indeed a helium 
gradient below the convective zone which is smoother than the gradient obtained 
with
 pure settling. Introduction of macroscopic motions in competition with the 
settling 
slightly smoothes down the helium gradient, and may rub out the peak, although 
some differences still remain between the models and the seismic inversion 
results (Brun, Turck-Chi\`eze, Zahn 1999, Richard, Th\'eado and Vauclair 2002).

Such motions are also needed to reproduce the observed abundances of light 
elements, namely a lithium depletion by about 140 compared
to the protosolar
value while beryllium is normal (Balachandran and Bell 1998).
Furthermore,  observations of the $^3$He/$^4$He ratio in the solar
wind and in the lunar rocks (Geiss and Gloecker 1998) show that this ratio may 
not
have increased by more than $\cong~10\%$ during the past 3 Gyr in the Sun.
While the occurrence of some mild mixing below
the solar convective zone is needed to explain
the lithium depletion and helps for the conciliation of the models with 
helioseismological constraints,
the $^3$He/$^4$He
observations put a strict constraint on its efficiency.
The effect of $\displaystyle \mu $-gradients on the mixing processes has to be 
invoked to explain 
these observations in a consistent way.

\subsection{Solar type stars}

Oscillations of solar 
type stars
have already been observed from ground based instruments (Bouchy and Carrier 2003).
Future space missions like Corot and Eddington will hopefully bring important new data.
We have begun to study the oscillations of 
stellar models including or not diffusion
or metal accretion, 
with the aim to test special signatures of these processes.
The models are iterated so that their external 
parameters ( $T_{eff}$, $L$, log $g$, metallicity)
are the same while their past history, and accordingly their present internal structure, are different.
Complete results will be presented elsewhre (Th\'eado et al 2004, 
Bazot and Vauclair 2004). Here we show some preliminary results : Figure 1 displays
evolutionary tracks for 1.2 M$_{\odot}$ stars, with and without diffusion,
iterated so that the outer stellar parameters remain quite similar (Castro 2003). Oscillation frequencies have
been computed for the models corresponding to the crossing point, which have the same luminosity and temperature
: log$(L/L_{\odot}) = 0.388$  and log$(T_{eff}) = 3.8$ .
Figure 2 displays, for the two models, the ``second
differences", namely $ \nu_{n+1}+\nu_{n-1}-2.\nu_n$. 
As pointed out by Gough 1989, these ``second differences" present characteristic 
oscillations due to the partial reflexion of the sound waves
on the regions of rapid variations of the sound velocity. Figure 2 shows clearly that they differ
in the two models. 
Detailed analysis of all these effects is presently underway.

\begin{figure}
\plotone{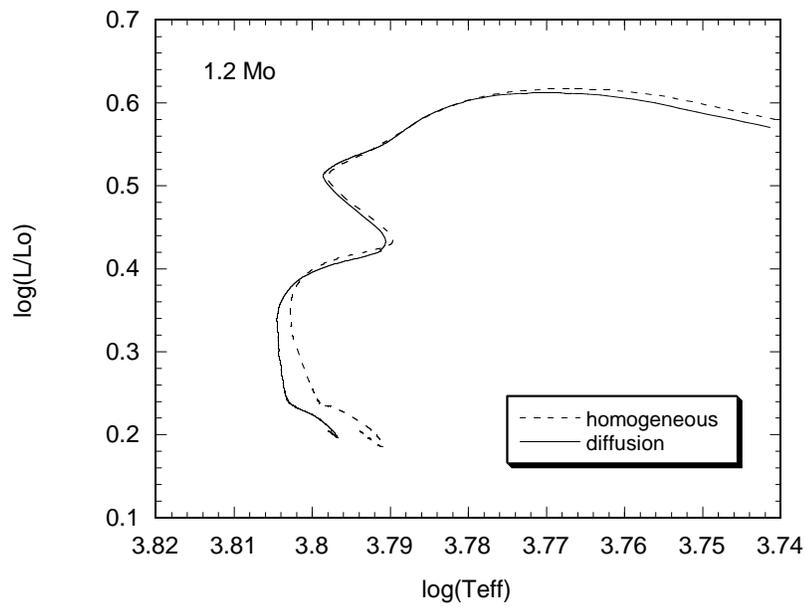}
\caption{Evolutionary tracks of two 1.2 M$_{\odot}$ stars ,with and without diffusion, iterated so that their outer (observable) parameters look the same; their
internal structure is different and this difference may be tested by studying the oscillation frequencies (after Castro 2003).   \label{fig1}}
\end{figure}

\begin{figure}
\plotone{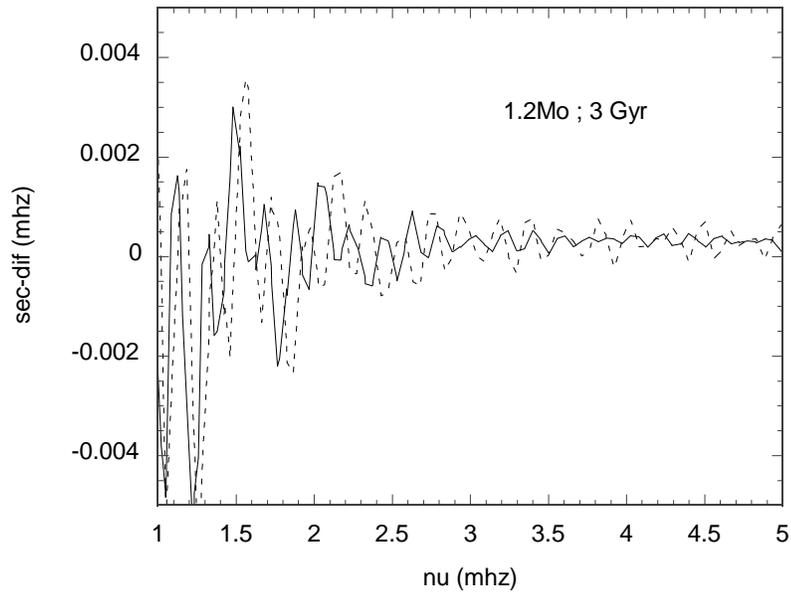}
\caption{Second differences (see text) for the two models corresponding to the crossing of the tracks (Figure 1); we can see that they have different variations, which is attributed to a different
internal structure; such effects will be precisely analysed in a forthcoming paper \label{fig2}}
\end{figure}

From these preliminary results, we may expect  
that the detailed observations and analysis
 of the oscillations of solar type stars will 
give clear signatures of 
diffusion processes in their interiors.

A fascinating new era is now opening in asteroseismology and stellar physics!

\end{document}